%% file: template.tex
\documentclass{webofc}

\usepackage[varg]{txfonts}   
\usepackage{hyperref}
\usepackage{url}
\usepackage{lineno}
\hypersetup{colorlinks=true,citecolor=blue,urlcolor=blue,linkcolor=blue}
\begin{document}

%
\input{commands.tex}
\title{Direct-photon measurement in small systems and thermal radiation from QGP with ALICE}
%
%

\author{\firstname{Jerome} \lastname{Jung for the ALICE Collaboration}\inst{1}\fnsep\thanks{\email{jung@ikf.uni-frankfurt.de}}
}

\institute{Johann Wolfgang Goethe-University Frankfurt am Main, Germany}

\abstract{

Electromagnetic probes are a unique tool for studying the space-time evolution of the hot and dense matter created in ultra-relativistic heavy-ion collisions. Dielectrons are emitted during the entire evolution of the medium created in such collisions, allowing the extraction of the real direct photon fraction at vanishing mass and providing access to thermal radiation from the early hot stages of the collision. The measurement of dielectron and direct-photon production in minimum-bias pp collisions serves as a crucial baseline for the studies in heavy-ion collisions, whereas pp collisions with high charged-particle multiplicities allow the search for the possible presence of QGP in small systems.

An overview of the final results of dielectron production in pp at $\sqrt{s}=13$ TeV and in central \PbPb collisions at \fivenn measured by ALICE is presented, together with their implications for the production of thermal radiation. Finally, an outlook on future measurements is given with the first performance studies from Run 3.

}
\maketitle

\section{Introduction}
\label{intro}

In central lead-lead (\PbPb) collisions at LHC energies, the energy densities are sufficient to create a hot and dense medium, the quark-gluon plasma (QGP).
Dielectrons (\ee pairs) are exceptionally well suited to study and constrain the space-time evolution of such collisions. They are produced at all stages of the collision and, due to their electromagnetic nature, leave the system with negligible final-state interaction, keeping the information of their production origins. The invariant mass (\mee) of dielectrons can be exploited to gain insight into the various stages of the collision since dielectrons of higher \mee must be produced in the earlier stages of the collision. 
At high invariant masses ($\mee > 3.1$~\GeVmass), dielectrons give insight into the initial stages dominated by hard scatterings and pre-equilibrium processes.
In the intermediate-mass range (IMR), i.e.~$1.2 < \mee < 2.6$~\GeVmass, the dielectron spectrum becomes sensitive to the thermal radiation of the partonic medium, thereby providing a measure of its temperature. Thermal radiation emitted from the hot hadronic phase via decays of $\rho$ mesons, whose spectral function is connected to chiral-symmetry restoration, is measurable in the mass region above the $\pi^0$ mass. Finally, at diminishing dielectron mass, the equivalence of virtual and real photons is exploited to extract the contribution of virtual direct photons. However, the extraction of the dielectron signal is obstructed by a significant background from combinatorial pairs and hadronic decays. Especially, the physical background of correlated semileptonic decays of heavy-flavour (HF) hadrons presents a major obstacle in the analysis and complicates the extraction of the QGP temperature in the IMR. Thus, it is crucial to control the HF contribution experimentally.\\
Measurements in minimum-bias (MB) proton-proton (pp) collisions provide an important vacuum baseline. For the study of heavy-flavour production, it represents a medium-free and therefore unmodified reference while the measurement of direct photons in such colliding systems tests the applicability of perturbative QCD (pQCD) calculations. In contrast, studies in high charged-particle multiplicity (HM) events can be utilized to search for the onset of thermal radiation in small systems.

\section{Results}

\begin{figure}
\centering
\includegraphics[width=3.8cm,clip]{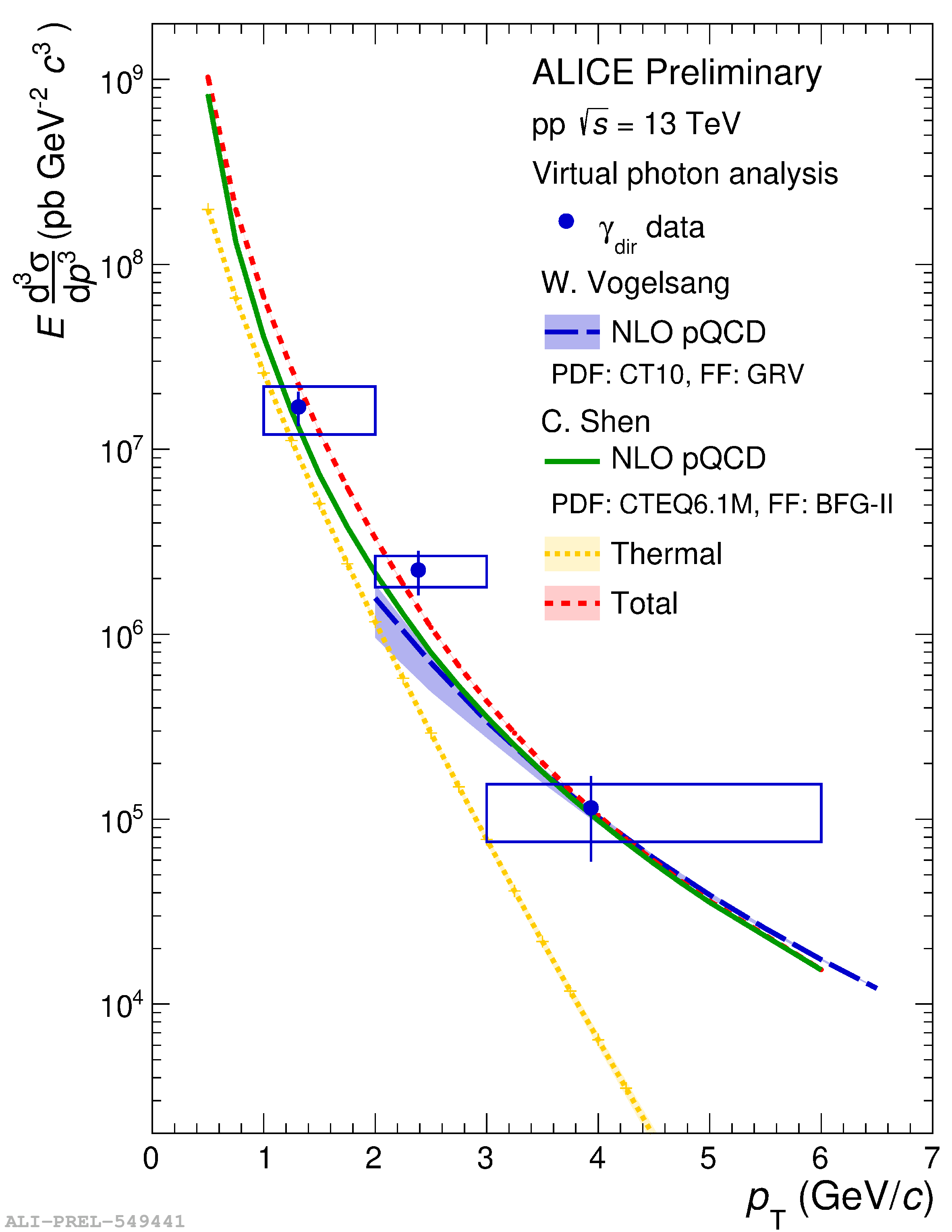}
\includegraphics[width=5.cm,clip]{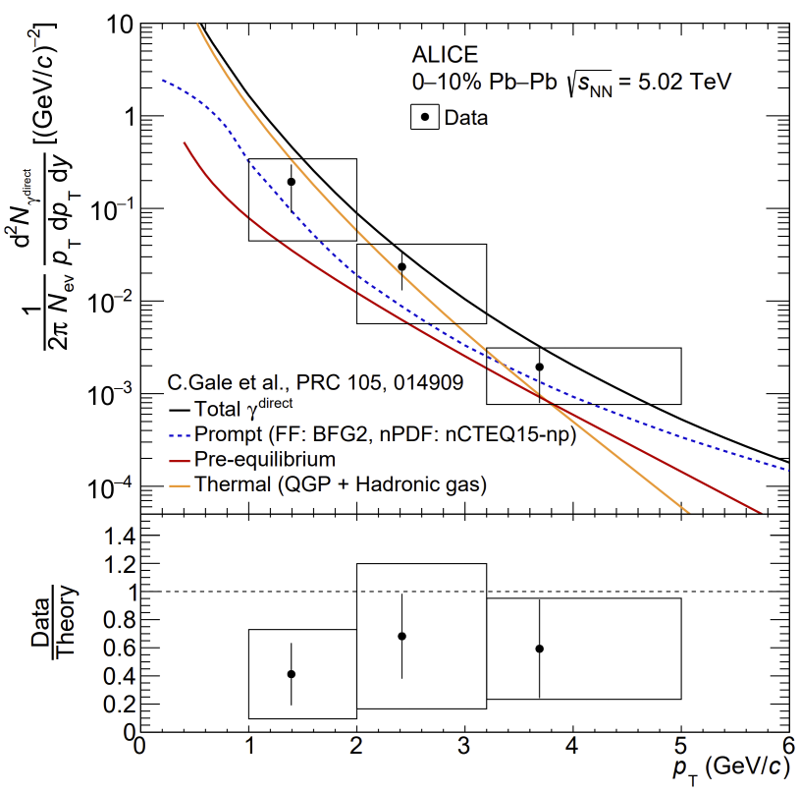}
\caption{Invariant yield of direct photons as a function of \pt measured in pp collisions at $\sqrt{s}=13$ TeV (left) and in the 10\% most-central \PbPb collisions at \fivenn (right), compared with different model predictions.}
\label{fig:DirectPhotons}       
\end{figure}

\begin{figure}
\centering
\includegraphics[width=5.cm,clip]{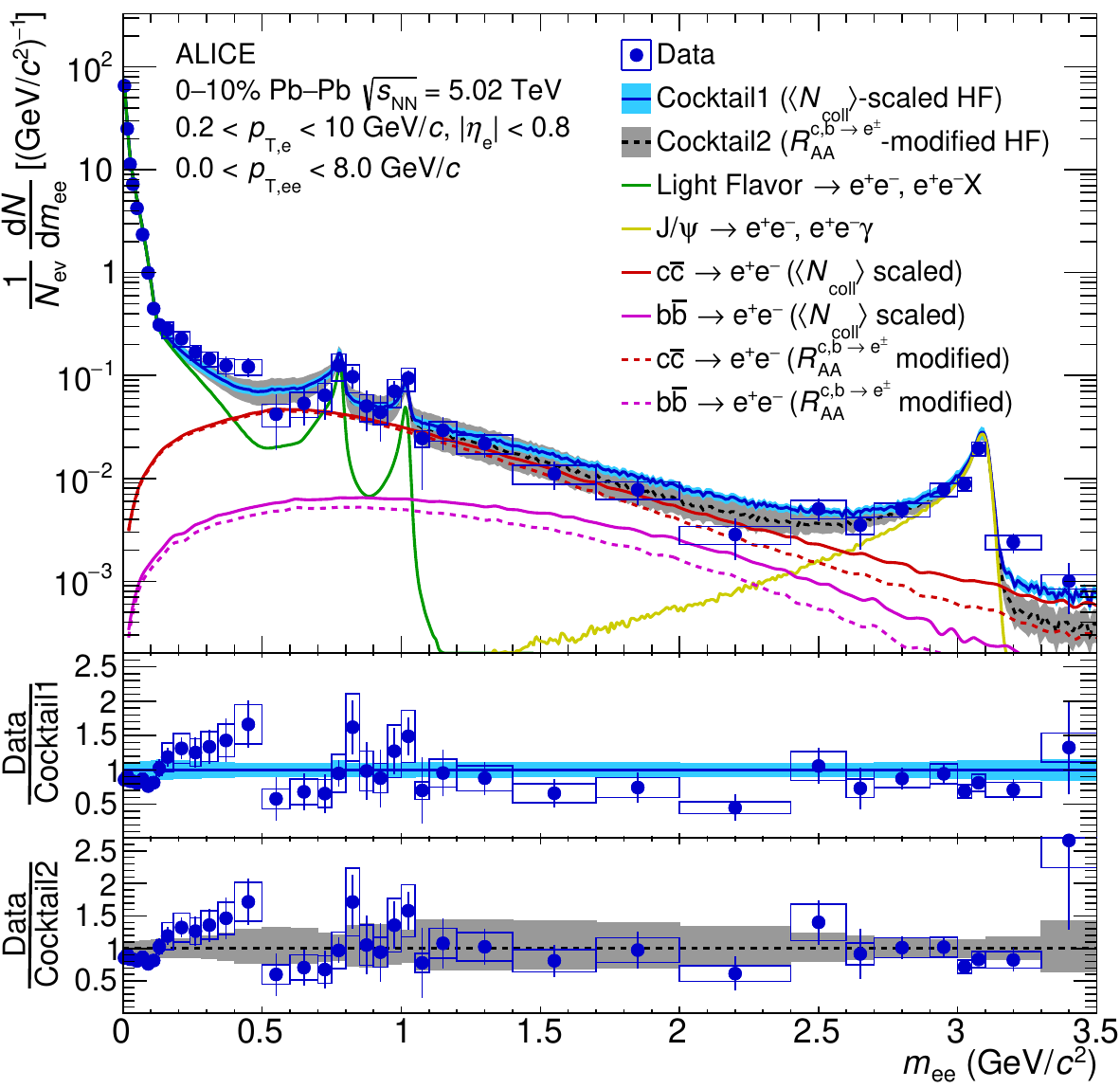}
\includegraphics[width=5.cm,clip]{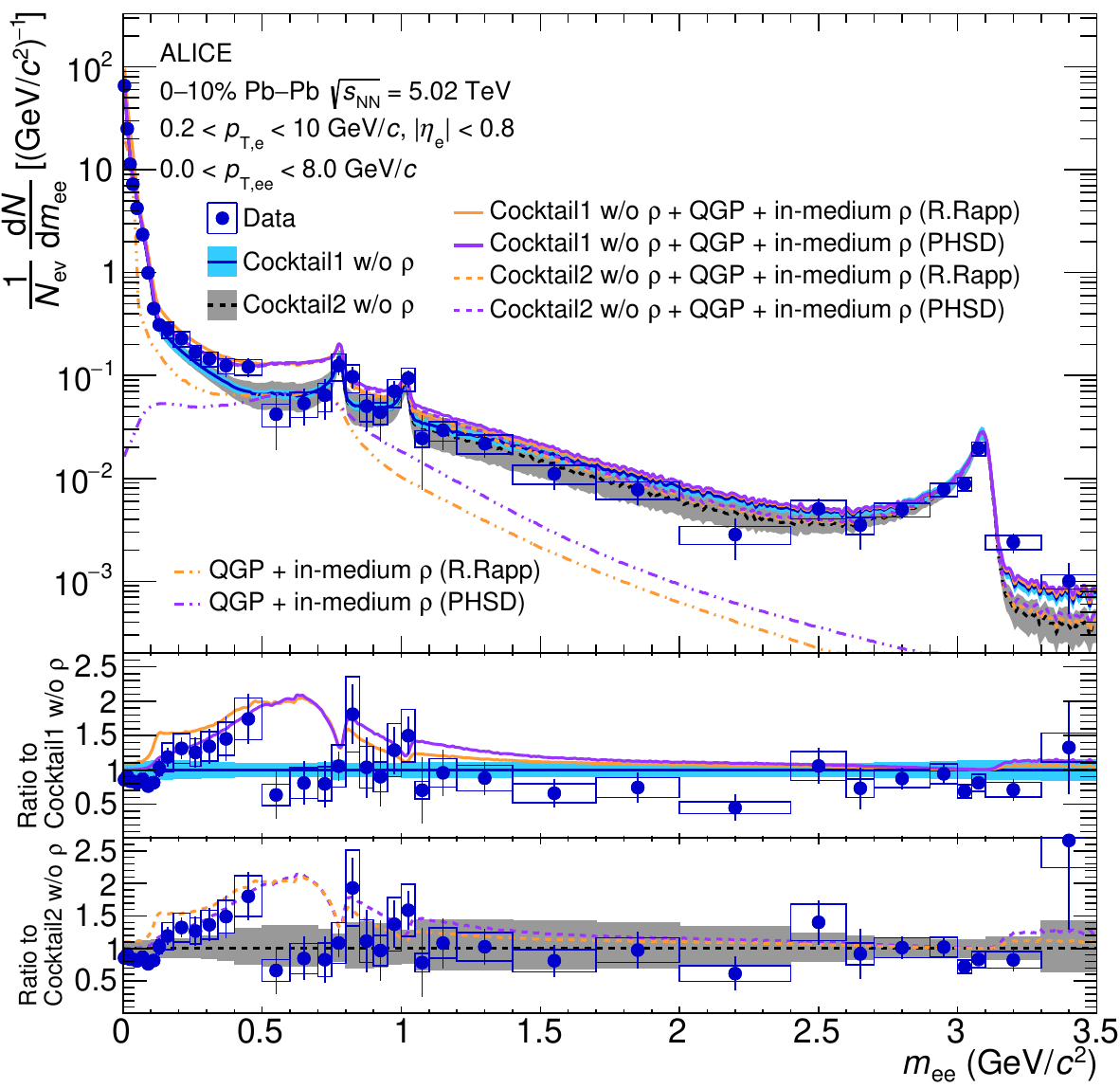}
\caption{Dielectron yield in the 10\% most central \PbPb collisions at \fivenn as a function of \mee, compared with two hadronic cocktails with different estimations for the correlated heavy-flavour contribution (left panel), and two predictions for thermal radiation from the medium~\cite{Rapp:2013nxa,Song:2018xca} (right panel).}
\label{fig:PbPbspectrum}       
\end{figure}

\begin{figure}
\centering
\includegraphics[width=5cm,clip]{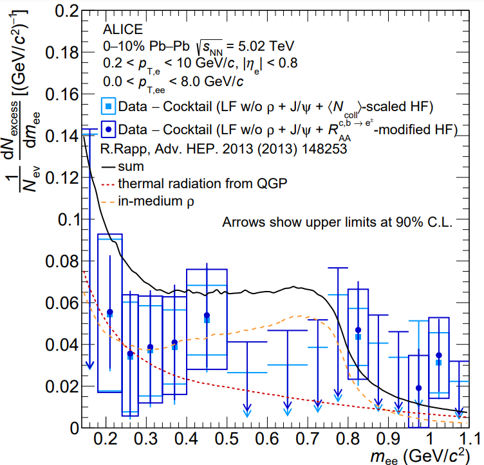}
\hspace{0.5cm} 
\includegraphics[width=5cm,clip]{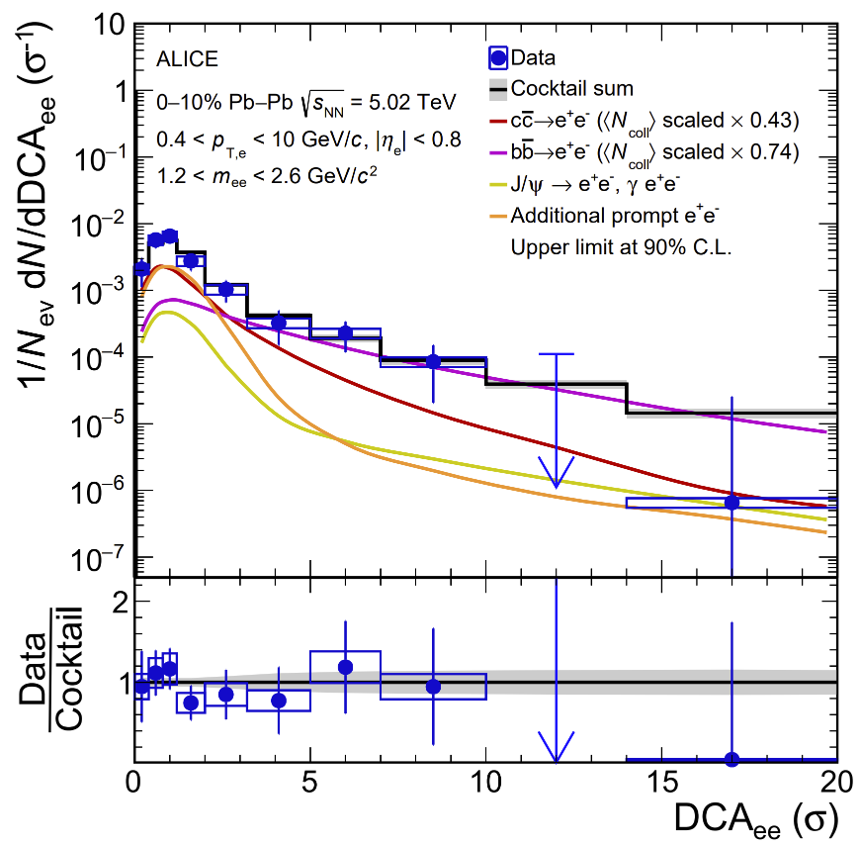}
\caption{Dielectron production in the 10\% most-central \PbPb collisions at \fivenn. Left: Excess yield after subtraction of known hadronic sources compared with predictions from the model of R. Rapp~\cite{Rapp:2013nxa}. Right: Template fit to the inclusive dielectron yield as a function of DCA$_{\rm ee}$ in the IMR.
}
\label{fig:PbPbResults}       
\end{figure}

The final results of the dielectron production in pp collisions at $\sqrt{s}=13$ TeV were obtained with the full Run 2 data set. Compared with the previous publication~\cite{Acharya_2019}, the number of analysed MB and HM events was increased by a factor of 3.8 and 4.4, respectively. Simultaneously, a more precise hadronic decay cocktail estimation has been constructed by updating the $\pi{^0}$ and $\eta$ meson contributions, making use of new independent neutral meson measurements in the same multiplicity intervals~\cite{alicecollaboration2024lightneutralmesonproductionpp}.
In MB collisions, the comparison of the inelastic dielectron cross section as a function of \mee with the hadronic expectation shows that the dielectron production (for \ptee > 1 GeV/$c$) can be well described by known hadronic sources. In the HM analysis, the hadronic cocktail is less constrained due to the unknown multiplicity dependence of the HF production, resulting in larger cocktail uncertainties and therefore allowing no observation of thermal radiation within uncertainties.
Applying the same method as in Ref. \cite{PHENIX:2008uif}, the ratio of direct-to-inclusive photon yields ($r$) is determined by fitting the dielectron spectrum in $0.14 < \mee < 0.4$~\GeVmass for different \ptee intervals. The extracted $r$ is then multiplied by an independently measured inclusive photon yield to construct a real direct-photon spectrum. The additional mass of dielectrons compared to real photons allows to suppress the large background from $\pi^{0}$ by fitting the spectra above the $\pi^{0}$ mass, reducing the systematic uncertainties.
\autoref{fig:DirectPhotons} (left) presents the first measurement of low \pt direct photons in small systems at LHC energies. The data is overlayed with two theoretical predictions: a pQCD calculation and a model containing an additional thermal radiation contribution. In both cases, the MB data is reproduced within the current experimental uncertainties. Comparing the HM results to the direct-photon yield measured in MB events, a significant increase is observed, which seems to be consistent with the relative increase in charge-particle multiplicity. A theoretical calculation of the photon production in HM events remains challenging due to the unknown scaling of the pQCD contribution with the increasing charge-particle multiplicity.
Following the same procedure, the first \pt-differential spectrum of direct photons at \fivenn was extracted~\cite{alicecollaboration2023dielectronproductioncentralpbpb} and is shown in \autoref{fig:DirectPhotons} (right). Here, the data are compared to a hybrid model~\cite{Gale:2021emg} that contains contributions from all stages of the collision. Within the current uncertainties, the direct-photon yield can be explained solely by the contribution of prompt photons. However, in this case, all central values are systematically larger than the pQCD baseline. In contrast, when the contributions of the pre-equilibrium phase and thermal radiation from the QGP and the hot hadronic phase are included, the data are overestimated by about 1$\sigma$. The \pt-integrated direct photon yield was studied as a function of the charged-particle multiplicity \dndeta. At RHIC energies, PHENIX reported previously an approximate power-law dependence of the measured direct-photon yield with \dndeta, independent on the collision energy or centrality.  Nevertheless, results from STAR and PHENIX disagree with decreasing \dndeta. In \PbPb collisions, ALICE measurements in the real-photon and virtual-photon channels are consistent with model predictions. The measurement obtained with virtual photons in pp collisions at $\sqrt{s} = 13$ TeV seems to align with the trend suggested by the larger systems and the STAR measurements, providing further input to constrain theoretical developments. However, the results at LHC energies are not sensitive enough yet to either confirm a universal scaling behaviour or to observe an onset of thermal radiation.
The measurement of dielectron production in the 10\% most-central \PbPb collisions at \fivenn~\cite{alicecollaboration2023dielectronproductioncentralpbpb} as a function of \mee is displayed in \autoref{fig:PbPbspectrum}. The result is compared to two hadronic cocktails, which differ in their estimation of the HF contribution. The first cocktail presents a vacuum baseline, ignoring any suppression or decorrelation effects introduced by the medium created in heavy-ion collisions. It is based on the HF measurements in pp collisions at the same energy~\cite{ALICE:2020mfy} scaled with the number of binary nucleon--nucleon collisions $\langle N_{\rm coll} \rangle$. This approach already provides a good description of the dielectron yield in the $\pi^0$ and $J/\psi$ mass regions but deviations in the IMR indicate a suppression of the HF contribution compared to pp collisions.
In the second cocktail, the HF modification is estimated based on the measurement of single HF decay electrons in \PbPb collisions~\cite{ALICE:2019nuy}. As cold (CNM) and hot nuclear matter (HNM) effects modify \ee pairs differently, they are disentangled using the measured $R^{{\rm c,b\to e^{\pm}}}_{\rm AA}$ and the one prediced with the EPS09 nuclear parton distribution functions~\cite{Eskola:2009uj}. The modifications are then propagated to \ee pairs using Monte Carlo (MC) simulations and a weighting procedure. Following this procedure, CNM and HNM effects seem to cancel out at low \mee and \ptee within large uncertainties, while at large \mee or \ptee the \ee pairs are more suppressed due to HNM effects. This approach leads to an overall better description of the data in the IMR but results in significantly larger cocktail uncertainties arising from the model assumptions and the uncertainties of its inputs that limit the interpretation of the data. 
Independent of the method used to estimate the heavy-flavour contribution, both cocktail comparisons show at low masses \mee ($0.18 < \mee < 0.5$~\GeVmass) a systematic deviation from the hadronic expectation with a significance of 1.53$\sigma$ (1.3$\sigma$). An excess of dielectrons in this region is expected due to thermally produced $\rho$ mesons.
Due to its short lifetime compared to the duration of the hot hadronic phase and its strong coupling to the $\pi^{+}\pi^{-}$ channel, the $\rho$ meson is likely to be regenerated in the hot hadronic phase, leading to a broadening of its in-medium spectral function. This additional source of dielectrons is not included in the hadronic decay cocktail. Therefore, two different models for the production of thermal radiation are considered. The expanding fireball model by R. Rapp~\cite{Rapp:2013nxa} estimates the thermal emission rate of dielectrons from the hadronic phase based on a hadronic many-body theory with in-medium modified $\rho$, and utilizes a lattice-QCD inspired approach for the equation of state in the QGP. The Parton-Hadron-String Dynamics (PHSD) model follows a transport model approach~\cite{Song:2018xca}, including in-medium modified electromagnetic spectral functions of low-mass vector mesons. 
The excess spectrum, obtained by subtracting all known hadronic sources of the cocktail, excluding the contribution of the $\rho$ meson, is shown in \autoref{fig:PbPbResults} (left). Both models are compatible with the data within experimental uncertainties. However, a tension between the model predictions and the data around $0.5 < \mee < 0.7$~\GeVmass can be observed where the models overestimate the data by 2.7$\sigma$ or 4.0$\sigma$ depending on the respective cocktail. To resolve this tension, future measurements will be required.\\ 
In the IMR, where the modified HF contribution dominates the spectrum, a cocktail-independent method is needed to achieve the necessary precision to measure the thermal radiation of the QGP. A promising approach for this is given by the distance-of-closest approach (DCA) analysis. It allows to separate prompt from non-prompt dielectron sources based on their delayed decay. Electrons and positrons produced at the primary vertex show a small DCA$_{\rm e}$ compared to the ones originating from the semileptonic decays of open-charm and open-beauty hadrons with their finite decay lengths of $c\tau_{\rm D} \approx 150$~$\mu{\rm m}$ and $c\tau_{\rm B} \approx 470$~$\mu{\rm m}$, respectively. 
To select the corresponding pairs, the DCA of the pair is defined by taking the quadratic mean of the DCA$_{\rm e}$ of both tracks: ${\rm DCA_{ee}} = \sqrt{ \frac{1}{2} [{({\rm DCA}_{xy,1}/\sigma_{xy,1})^{2}+({\rm DCA}_{xy,2}/\sigma_{xy,2})^{2}}]}$, where DCA$_{xy,i}$ is the DCA of the electron $i$ in the transverse plane and $\sigma_{xy,i}$ represents its resolution.
The method is first established in the $J/\psi$-mass range, where the spectrum is expected to be dominated by contributions of prompt and non-prompt $J/\psi$ that are well constrained by independent ALICE measurements. 
For each dielectron source, DCA$_{\rm ee}$ templates are extracted from full MC simulations of the ALICE detector and scaled to the expectation of the hadronic cocktail. The comparison to the data shows a good agreement within experimental uncertainties, thereby validating both the approach and the DCA description in the MC simulation.
In the next step, this procedure is applied in the IMR. 
Here, the templates are fitted to the spectrum in a two-stage process. In a first step, a template fit is performed at high \ptee, where beauty is expected to dominate the spectrum, to constrain its contribution. This leads to a suppression factor of $0.74 \pm 0.24 \ {\rm (stat.)} \pm 0.12 \ {\rm (syst.)}$ for beauty with respect to $N_{\rm coll}$-scaling. In a second step, the remaining contributions of charm and prompt sources are determined via a simultaneous fit of their templates, presented in \autoref{fig:PbPbResults} (right). The data are consistent with a charm suppression by a factor $0.43 \pm 0.4 \ {\rm (stat.)} \pm 0.22 \ {\rm (syst.)}$ with respect to $N_{\rm coll}$-scaling and an additional prompt component that is a factor $2.64 \pm 3.18 \ {\rm (stat.)} \pm 0.29 \ {\rm (syst.)}$ larger than the predictions from R. Rapp~\cite{Rapp:2013nxa}. Unlike the cocktail-based method, this approach is no longer limited by its systematic uncertainties, and a larger data sample will enable the extraction of thermal dielectrons in the IMR in future measurements.

\section{Outlook}
\begin{figure}
\centering
\includegraphics[width=5.cm,clip]{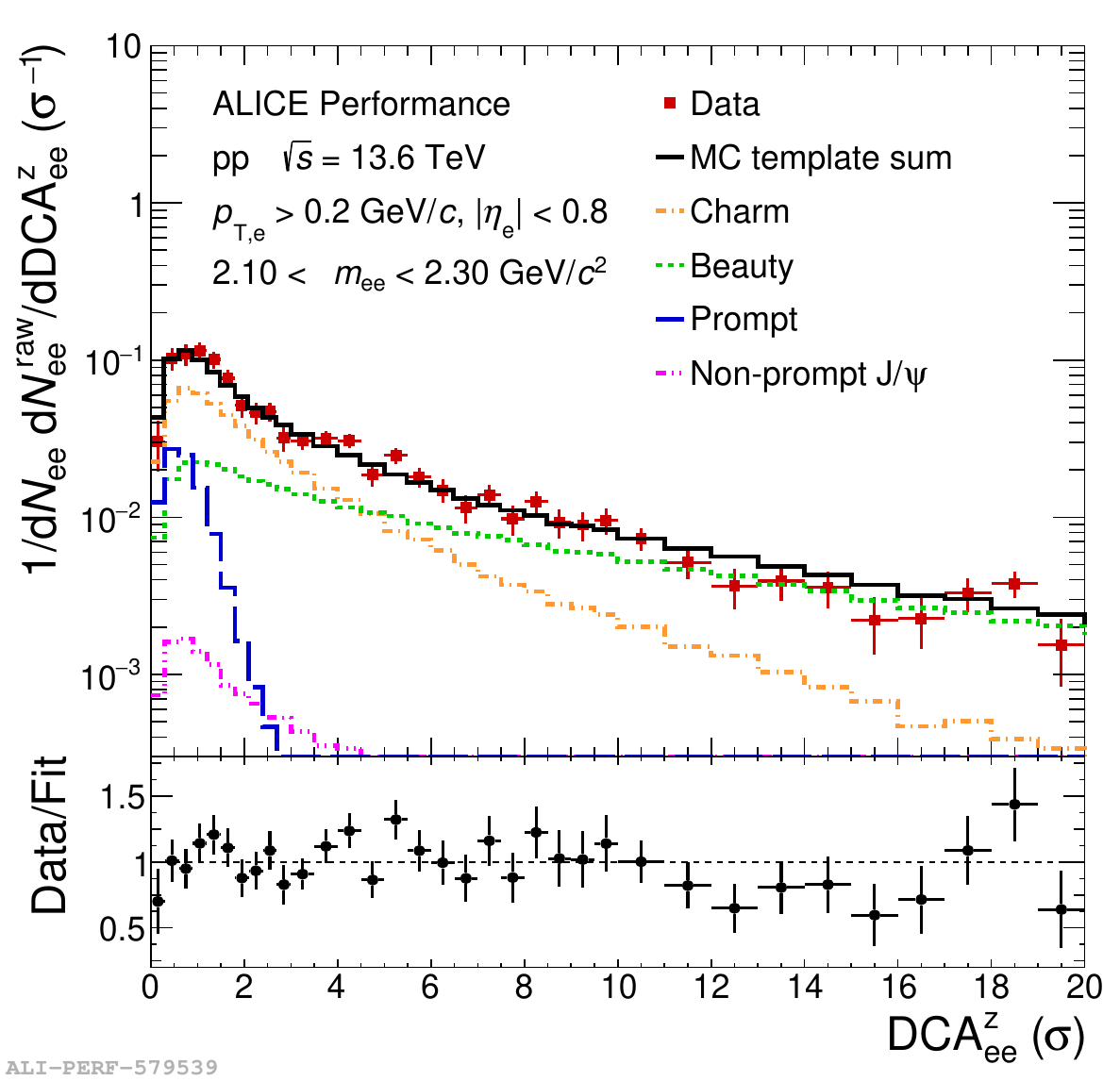}
\includegraphics[width=5.cm,clip]{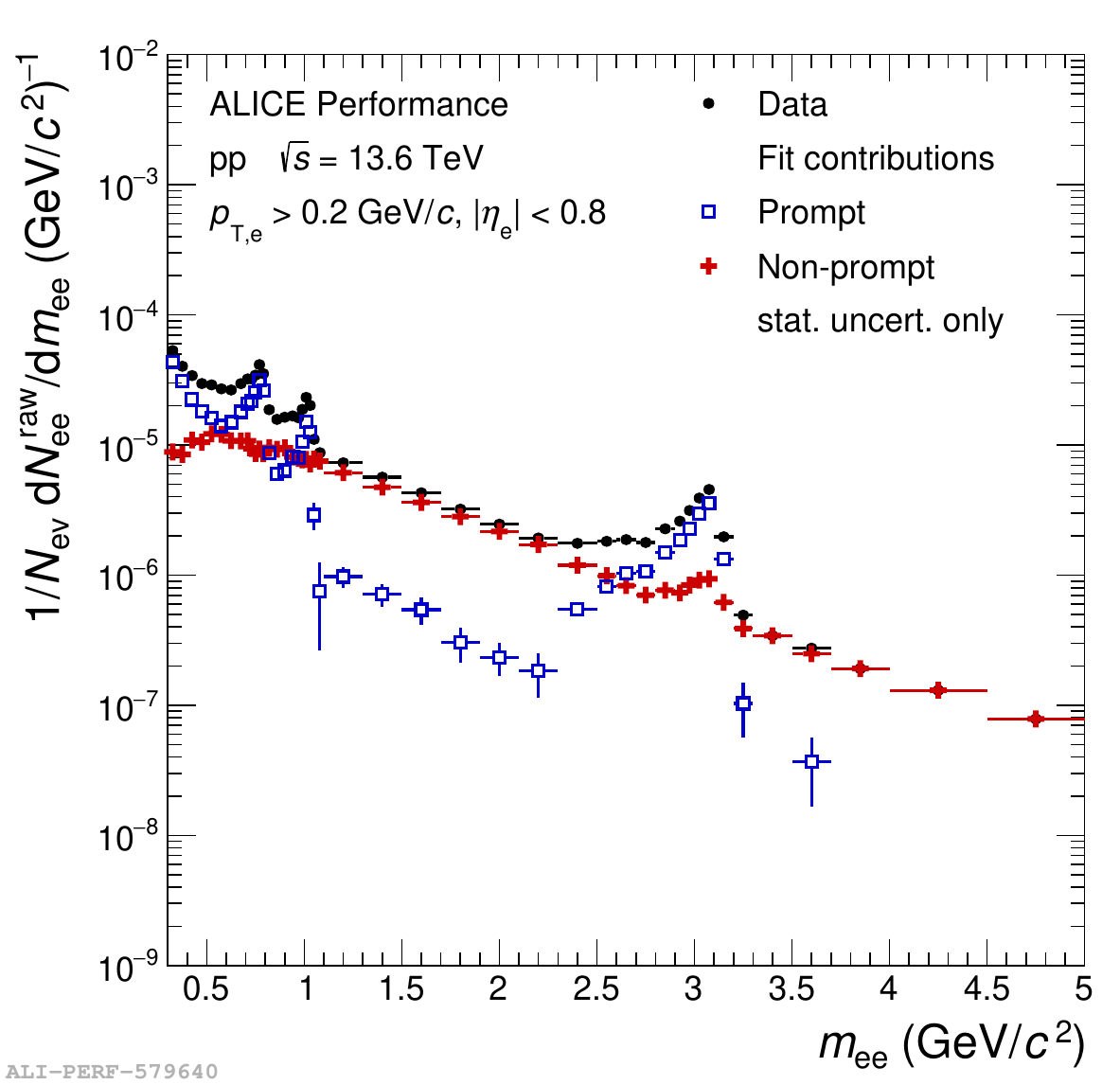}
\caption{Raw dielectron yield as a function of DCA$_{\rm ee}^{z}$ (DCA$_{\rm ee}$ in the beam direction) measured in pp collisions at $\sqrt{s}=13.6$ TeV in the IMR fitted with different templates (left). Corresponding invariant-mass spectra unfolded for prompt and non-prompt sources based on DCA$_{\rm ee}^{z}$ template fits.}
\label{fig:Run3ppDCA}       
\end{figure}

During the Long Shutdown 2, the ALICE apparatus underwent a major detector upgrade~\cite{Acharya_2024}. The new GEM-based readout system of the TPC allows a significant increase of the readout rate by a factor of 1000 in pp and 100 in \PbPb. In addition, the upgrade of the ITS to the CMOS MAPS technology improves the single-track DCA resolution by a factor of 2 to 5. 
In Runs 3 and 4, pp collisions at $\sqrt{s}=13.6$ TeV are expected to be recorded with an integrated luminosity of 200 $pb^{-1}$. A large MB data set of 0.97 $pb^{-1}$ was already collected in 2022, increasing the statistics compared to Run 2 by over a factor of 30.
This dramatic increase in statistics enables template fits to the raw DCA$_{\rm ee}$ spectra in small \mee intervals, as presented in \autoref{fig:Run3ppDCA} (left). The scaled templates of charm, beauty, non-prompt $J/\psi$, and prompt sources give a good description of the data over the full DCA$_{\rm ee}$ range. Repeating this process over the full mass range allows unfolding the invariant mass spectrum depending on the specific decay kinematics, separating prompt and non-prompt sources experimentally. The result can be seen in \autoref{fig:Run3ppDCA} (right). 
The non-prompt contribution shows a smooth distribution following the heavy-flavour continuum with the exception of a small peak due to the non-prompt $J/\psi$ originating from beauty feed-down processes. In contrast, the prompt distribution displays the resonance structures of the pseudoscalar and vector mesons at lower masses, and a significant $J/\psi$ peak at higher masses, while in the IMR a non-vanishing prompt contribution is preferred by the fit. Overall, this first performance study illustrates the capabilities of the new detector and serves as an important baseline for future \PbPb measurements.\\
An integrated luminosity of 13 $nb^{-1}$ is planned for \PbPb collision at \snn = 5.36 TeV. During 2023, already 1.5 $nb^{-1}$ were recorded during the first heavy-ion data taking campaign. A first look at the raw-signal extraction as a function of \mee in 10-90$\%$ \PbPb collisions with an electron selection of $\pte>0.4$ GeV/$c$ is presented. It shows the characteristic signatures of the $\pi^{0}$ and $J/\psi$ peaks, and the signal-to-background ratio in the IMR is compatible with Run 2 in central collisions. In the next steps, the analysis will be extended to include most-central events as well as the topological separation via DCA$_{\rm ee}$.

\section{Conclusion}
\label{sec:Conclusion}
The dielectron analyses performed on the Run 2 data are getting finalized.
Results obtained with the full Run 2 dataset show significantly reduced statistical and systematic uncertainties compared to the previous publication in pp collisions at 13 TeV, allowing an extraction of the direct-photon fraction in MB and HM events. In central \PbPb collisions at \snn = 5.02 TeV, a first measurement of a direct-photon yield is performed and a first DCA$_{\rm ee}$ analysis to separate thermal radiation from the HF background in \PbPb collisions is established.
However, the uncertainties of the Run 2 data do not allow the extraction of a significant thermal signal. First performance studies with the upgraded ALICE detector have been performed in pp and \PbPb collisions. The data samples to be collected during the LHC Run 3 and 4 will significantly reduce the statistical uncertainties~\cite{Acharya_2024}. Thus, an extraction of QGP radiation in the IMR is expected to be possible in Run 3 and 4.

\bibliography{template.bib}

\end{document}

%% file: commands.tex
%

\newcommand{\pp}           {pp\xspace}
\newcommand{\ppbar}        {\mbox{$\mathrm {p\overline{p}}$}\xspace}
\newcommand{\XeXe}         {\mbox{Xe--Xe}\xspace}
\newcommand{\PbPb}         {\mbox{Pb--Pb}\xspace}
\newcommand{\pA}           {\mbox{pA}\xspace}
\newcommand{\pPb}          {\mbox{p--Pb}\xspace}
\newcommand{\AuAu}         {\mbox{Au--Au}\xspace}
\newcommand{\dAu}          {\mbox{d--Au}\xspace}

\newcommand{\mee}         {\ensuremath{m_{\rm ee}}\xspace}
\newcommand{\ptee}         {\ensuremath{p_{\rm T,ee}}\xspace}
\newcommand{\ptll}         {\ensuremath{p_{\rm T,ll}}\xspace}
\newcommand{\pteesquare}    {\ensuremath{p_{\rm T,ee}^{\rm 2}}\xspace}
\newcommand{\meanpteesquare}       {$\langle p_{\rm T,ee}^{\rm 2}\rangle$\xspace}
\newcommand{\s}            {\ensuremath{\sqrt{s}}\xspace}
\newcommand{\snn}          {\ensuremath{\sqrt{s_{\mathrm{NN}}}}\xspace}
\newcommand{\pt}           {\ensuremath{p_{\rm T}}\xspace}
\newcommand{\pte}           {\ensuremath{p_{\rm T,e}}\xspace}
\newcommand{\etae}           {\ensuremath{\eta_{\rm e}}\xspace}
\newcommand{\meanpt}       {$\langle p_{\mathrm{T}}\rangle$\xspace}
\newcommand{\ycms}         {\ensuremath{y_{\rm CMS}}\xspace}
\newcommand{\ylab}         {\ensuremath{y_{\rm lab}}\xspace}
\newcommand{\etarange}[1]  {\mbox{$| \eta |\,<\,#1$}}
\newcommand{\etarangee}[1]  {\mbox{$| \eta_{\rm e} |<#1$}}
\newcommand{\yrange}[1]    {\mbox{$\left | y \right |~<~#1$}}
\newcommand{\dndy}         {\ensuremath{\mathrm{d}N_\mathrm{ch}/\mathrm{d}y}\xspace}
\newcommand{\dndeta}       {\ensuremath{\mathrm{d}N_\mathrm{ch}/\mathrm{d}\eta}\xspace}
\newcommand{\avdndeta}     {\ensuremath{\langle\dndeta\rangle}\xspace}
\newcommand{\dNdy}         {\ensuremath{\mathrm{d}N_\mathrm{ch}/\mathrm{d}y}\xspace}
\newcommand{\Npart}        {\ensuremath{N_\mathrm{part}}\xspace}
\newcommand{\Ncoll}        {\ensuremath{N_\mathrm{coll}}\xspace}
\newcommand{\dEdx}         {\ensuremath{\textrm{d}E/\textrm{d}x}\xspace}
\newcommand{\RpPb}         {\ensuremath{R_{\rm pPb}}\xspace}

\newcommand{\nineH}        {$\sqrt{s}=0.9$~Te\kern-.1emV\xspace}
\newcommand{\seven}        {$\sqrt{s}=7$~Te\kern-.1emV\xspace}
\newcommand{\twoH}         {$\sqrt{s}=0.2$~Te\kern-.1emV\xspace}
\newcommand{\twoHnn}         {$\sqrt{s_{\mathrm{NN}}}=0.2$~Te\kern-.1emV\xspace}
\newcommand{\twosevensix}  {$\sqrt{s}=2.76$~Te\kern-.1emV\xspace}
\newcommand{\five}         {$\sqrt{s}=5.02$~Te\kern-.1emV\xspace}
\newcommand{\twosevensixnn}{$\sqrt{s_{\mathrm{NN}}}=2.76$~Te\kern-.1emV\xspace}
\newcommand{\fivenn}       {$\sqrt{s_{\mathrm{NN}}}=5.02$~Te\kern-.1emV\xspace}
\newcommand{\LT}           {L{\'e}vy-Tsallis\xspace}
\newcommand{\GeVc}         {Ge\kern-.1emV$/c$\xspace}
\newcommand{\MeVc}         {Me\kern-.1emV$/c$\xspace}
\newcommand{\TeV}          {Te\kern-.1emV\xspace}
\newcommand{\GeV}          {Ge\kern-.1emV\xspace}
\newcommand{\MeV}          {Me\kern-.1emV\xspace}
\newcommand{\GeVmass}      {Ge\kern-.1emV$/c^2$\xspace}
\newcommand{\MeVmass}      {Me\kern-.2emV$/c^2$\xspace}
\newcommand{\lumi}         {\ensuremath{\mathcal{L}}\xspace}

\newcommand{\ITS}          {\rm{ITS}\xspace}
\newcommand{\TOF}          {\rm{TOF}\xspace}
\newcommand{\ZDC}          {\rm{ZDC}\xspace}
\newcommand{\ZDCs}         {\rm{ZDCs}\xspace}
\newcommand{\ZNA}          {\rm{ZNA}\xspace}
\newcommand{\ZNC}          {\rm{ZNC}\xspace}
\newcommand{\SPD}          {\rm{SPD}\xspace}
\newcommand{\SDD}          {\rm{SDD}\xspace}
\newcommand{\SSD}          {\rm{SSD}\xspace}
\newcommand{\TPC}          {\rm{TPC}\xspace}
\newcommand{\TRD}          {\rm{TRD}\xspace}
\newcommand{\VZERO}        {\rm{V0}\xspace}
\newcommand{\VZEROA}       {\rm{V0A}\xspace}
\newcommand{\VZEROC}       {\rm{V0C}\xspace}
\newcommand{\Vdecay} 	   {\ensuremath{V^{0}}\xspace}

\newcommand{\ee}           {\ensuremath{{\rm e}^{+}{\rm e}^{-}}\xspace} 
\newcommand{\pip}          {\ensuremath{\pi^{+}}\xspace}
\newcommand{\pim}          {\ensuremath{\pi^{-}}\xspace}
\newcommand{\kap}          {\ensuremath{\rm{K}^{+}}\xspace}
\newcommand{\kam}          {\ensuremath{\rm{K}^{-}}\xspace}
\newcommand{\pbar}         {\ensuremath{\rm\overline{p}}\xspace}
\newcommand{\kzero}        {\ensuremath{{\rm K}^{0}_{\rm{S}}}\xspace}
\newcommand{\lmb}          {\ensuremath{\Lambda}\xspace}
\newcommand{\almb}         {\ensuremath{\overline{\Lambda}}\xspace}
\newcommand{\Om}           {\ensuremath{\Omega^-}\xspace}
\newcommand{\Mo}           {\ensuremath{\overline{\Omega}^+}\xspace}
\newcommand{\X}            {\ensuremath{\Xi^-}\xspace}
\newcommand{\Ix}           {\ensuremath{\overline{\Xi}^+}\xspace}
\newcommand{\Xis}          {\ensuremath{\Xi^{\pm}}\xspace}
\newcommand{\Oms}          {\ensuremath{\Omega^{\pm}}\xspace}
\newcommand{\degree}       {\ensuremath{^{\rm o}}\xspace}